\documentclass[preprint,showpacs,preprintnumbers,amsmath,amssymb]{revtex4}
\usepackage{graphicx}
\usepackage{dcolumn}
\usepackage{bm}

\begin{document}

\title{Stability and plasticity of silicon nanowires: the role of
wire perimeter}

\author{J. F. Justo$^{\rm (1)}$, R. D. Menezes$^{\rm (1)}$,  
and L. V. C. Assali$^{\rm (2)}$}

\affiliation{$^{\rm (1)}$ Escola Polit\'ecnica, 
Universidade de S\~ao Paulo,
CP 61548, CEP 05424-970, S\~ao Paulo, SP, Brazil}

\affiliation{$^{\rm (2)}$ Instituto de F\'{\i}sica, 
Universidade de S\~ao Paulo,
CP 66318, CEP 05315-970, S\~ao Paulo, SP, Brazil}


\begin{abstract}
We investigated the properties of stability and plasticity 
of silicon nanowires using molecular dynamics simulations. We considered 
nanowires with $\langle 100 \rangle$, $\langle 110 \rangle$ and 
$\langle 112 \rangle$ growth directions with several diameters and 
surface facet configurations. We found that the wire perimeter, 
and not the wire diameter, is the meaningful dimensional parameter. 
As a result, the surface facets play a central role on the nanowire 
energy, that follows a universal scaling law. Additionally, we have 
computed the response of a silicon nanowire to external load. 
The results were compared to available experimental and 
{\it ab initio} data.
\end{abstract}

\maketitle

\section{Introduction}

Silicon nanowires (SiNWs) belong to a unique class of 
semiconductor nanowires, since they may allow, in a near future, 
integration within conventional silicon-based device technology. In fact, 
the experimental realization of SiNW electronic devices has been 
achieved very recently \cite{cui}. Silicon nanowires have been 
synthesized by several 
methods \cite{morales,holmes}, generating wires in a wide range of 
diameters and crystallographic growth directions. Although the resulting 
wires are generally covered by oxide layers, complete removal of 
those oxide layers has been achieved \cite{ma}, leading 
to SiNWs with hydrogen passivated surfaces. SiNWs with reconstructed 
surfaces have also been produced \cite{marsen}.

SiNWs have been grown along several crystallographic 
directions: Holmes {\it et al.} reported SiNWs along the 
$\langle 100 \rangle$ and $\langle 110 \rangle$ directions with 
diameters of about 4 nm \cite{holmes}. Ma {\it et al.} reported 
ultra-thin SiNWs along the $\langle 100 \rangle$ directions with 
diameters as low as 1.3 nm \cite{ma}. Wu {\it et al.} reported SiNWs 
along $\langle 110 \rangle$, $\langle 111 \rangle$, and 
$\langle 112 \rangle$ lattice directions \cite{wu}. High resolution
electron microscopy experiments have shown that the resulting SiNWs 
carry cores with monocrystalline bulk structures \cite{ma,wu}.

Here, we carried a theoretical investigation 
on the stability of clean surface SiNWs along three growth directions 
and with a wide range of surface facet families. 
We found that the nanowire perimeter, and not its 
diameter, is the representative dimensional parameter to compute the
wire scaling effects. The nanowire energy, for a certain growth direction, 
follows a universal scaling law with relation to the inverse of the 
nanowire perimeter. Additionally, we 
investigated the behavior of SiNWs under tensile deformation and compared 
the results with recent theoretical \cite{menon} and experimental 
data \cite{kizuka}.

\section{methodology}

Theoretical investigations, based on quantum mechanical calculations, 
have computed the properties of stable 
ultra-thin SiNWs \cite{li,rurali,kagimura}. However, such calculations 
are very expensive and investigations have been restricted to systems 
involving only a few hundred atoms and at an equilibrium state. Size-dependent 
and thermodynamical properties of nanowires are still unattainable to 
such methods. Although empirical methods carry a 
considerable simplification of the underlying atomistic processes, 
they still represent an alternative to access those important 
nanowire properties \cite{menon,pono}.

Our simulations were performed using molecular dynamics, in which the 
Si-Si interactions were described by the Environment Dependent Interatomic
Potential (EDIP) 
model \cite{justo1,justo2}. Some wire configurations were also computed  
using the Stillinger-Weber (SW) \cite{sw} and 
Tersoff \cite{tersoff} potentials, but the final results showed that the
EDIP model provided the best description of silicon nanowires. 
The EDIP model \cite{justo1,justo2} includes environment terms that
capture the essential elements of Si-Si interactions in a wide range of 
atomic coordinations. That 
model represented a substantial improvement, as compared
to previous models \cite{sw,tersoff}, on the
description of several important materials properties, such as 
disordered phases \cite{brambilla,mousseau}, 
surfaces \cite{justo2,sinno}, point \cite{colombo} and
line \cite{justo3,justo4} defects, and
bulk elastic properties \cite{porter}. 
Therefore, this model appears to be an appropriate choice to 
investigate the stability and plasticity of SiNWs.

\begin{figure}
\centering{
\includegraphics[width=160mm]{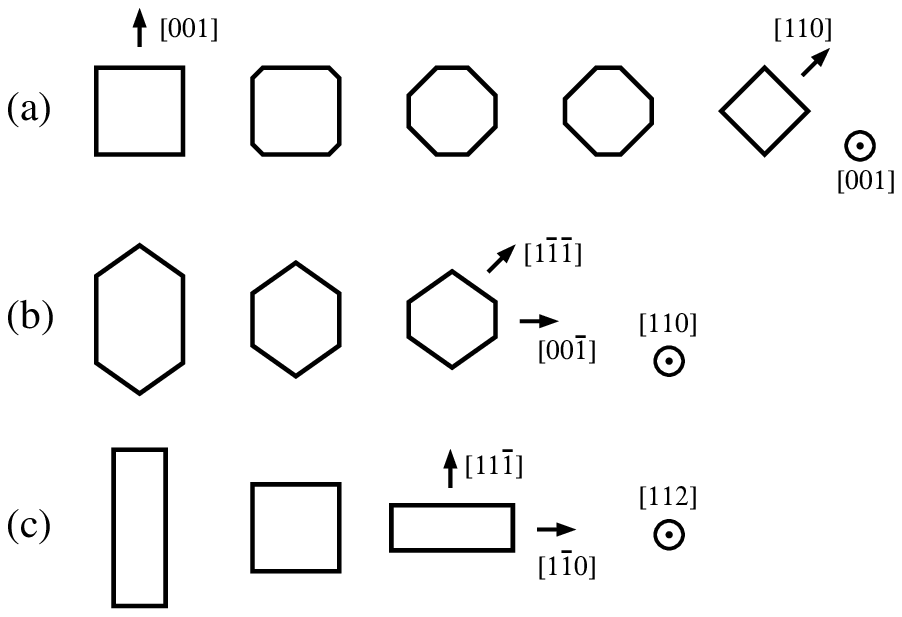}
\caption{SiNW configurations for different 
families in (a) $\langle 100\rangle$,  (b) $\langle 110\rangle$,
and (c)  $\langle 112\rangle$ growth directions. For  
$\langle 100\rangle$ wires, we considered configurations ranging from 
pure $\{100\}$ surfaces to pure $\{110\}$ surfaces. For  
$\langle 110\rangle$ wires, we considered those ranging 
from prevailing $\{100\}$ surfaces to prevailing $\{111\}$ ones. 
For $\langle 112\rangle$ wires, we considered  
those ranging from prevailing $\{110\}$ surfaces to 
prevailing $\{111\}$ ones.}
\label{fig1}
}
\end{figure}

Depending on the wire configuration, the simulations involved up to 
40,000 atoms. The wire stable configurations were computed by 
molecular dynamics, following an annealing procedure. 
The simulation of each wire started with a reasonably high temperature 
(600 K), being slowly cooled down to a few kelvins, 
when the statistical properties were computed. 
We investigated the stability of SiNWs for  
$\langle 100\rangle$, $\langle 110\rangle$, and
$\langle 112\rangle$ growth directions. Figure \ref{fig1} shows a 
cross-sectional representation of the wires investigated here. 
For $\langle 100\rangle$ wires (fig. \ref{fig1}a), we 
considered five families with different facets, ranging from wires with 
pure $\{100\}$ surfaces to wires with pure $\{110\}$ surfaces. 
For $\langle 110\rangle$ wires (fig. \ref{fig1}b), 
we considered wire families ranging from prevailing $\{100\}$ to 
prevailing $\{111\}$ surfaces. Due to geometrical reasons, there is 
no $\langle 110\rangle$ wire comprising only $\{100\}$ or $\{111\}$ 
pure surfaces. For $\langle 112\rangle$ wires (fig. \ref{fig1}c), 
we considered wire families  
ranging from wires with prevailing $\{111\}$ surfaces to 
wires with prevailing $\{110\}$ surfaces. The facet configurations for 
those growth directions were consistent with images of 
$\langle 110\rangle$ and $\langle 112\rangle$ SiNWs \cite{ma}. 
Specifically in the case of the $\langle 112\rangle$ wires, the images 
revealed that the wires contained $\{111\}$ surfaces \cite{ma}. Additionally,
the image intensity along the wire border suggested that those borders 
should be 
sharp. Considering those two elements, and that crystalline silicon $\{110\}$ 
surfaces have low energies, we assumed that the $\langle 112\rangle$ 
wires comprised only $\{111\}$ and $\{110\}$ surfaces.

\section{Results}

The scaling properties of nanowires have been
described as a function of their 
diameters \cite{rurali,kagimura,kizuka}. However, defining a nanowire 
diameter has been a challenging task, since nanostructures 
based on covalent bonding 
generally have facets and do not have a single diameter. Authors either 
avoid defining such a parameter \cite{pono,rurali} or describe the wire 
representative dimension as its smallest diameter, taken from images of 
the wire cross-section \cite{kizuka}. Others take the diameter of 
the smallest cylinder that contains the wire \cite{ma,kagimura}. 
Ultimately, it is assumed that the nanowire has a prevailing cylindrical 
shape \cite{kagimura}. For large diameters, 
properties are reasonably well described using any of those assumptions, 
but not for thinner wires. Considering the SiNW faceting, 
we find that the wire perimeter ($P$), and not the
wire diameter, 
provides an appropriate description of the nanowire scaling properties. 
The wire perimeter comprises the sum of the length of each facet ($f_i$) of 
the wire ($P=\sum f_i$). The surface size of each facet is 
determined by $f_i \times L$ ($L$ is the wire length) and the total wire 
surface is given by $P \times L$. As a result, a wire
scaling law described in terms of its perimeter is equivalent to 
a law in terms of its total surface. Considering
that in a nanowire, the surface/volume ratio is very large, it is
reasonable to consider that scaling laws should be described
in terms of the wire surface.

The nanowire total energy could be described 
by an analytical model, as recently discussed in ref. \cite{yako}.  
This energy ($F$) comprises three elements: 
a bulk ($E_b$), a surface ($E_s$), and an edge ($E_e$) term,
\begin{equation}
F = E_{e} +  E_{s} +  E_{b} \ ,
\end{equation}
where the surface term is given by the contribution of all wire facets:
\begin{equation}
 E_{s} = \sum_i \gamma_i s_i 
\end{equation}
where $\gamma_i$ is the surface energy of facet $i$, and 
$s_i$ is the number of unit cells in the surface \cite{yako}.

The nanowire energy lies between two limits:
\begin{equation}
E_{e} + \gamma_{min}  \sum_i  s_i < (F - E_{b}) <
E_{e} + \gamma_{max} \sum_i  s_i  
\end{equation}
where $\gamma_{min}$ and $\gamma_{max}$ represent 
respectively the minimum and
maximum values for the surface energy.
Now, dividing all the terms by the number of atoms, 
$N  \propto $ P$^{2}$, per unit
length, one gets the following relation 
for the wire energy per atom (within some constant scale):
\begin{equation}
E_{e}P^{-2} + \gamma_{min}P^{-1} < (F - E_{b})/N <
E_{e}P^{-2} + \gamma_{max}P^{-1} . 
\label{eq10}
\end{equation}

Equation \ref{eq10} gives the limits for wire energy per atom, 
$(F - E_{b})/N$,
in terms of the wire perimeter. 
For large perimeters, the edge effects could be neglected, and
the wire energy should have a linear relation with  $P^{-1}$,
and lie between two limiting cases, that are controlled by
$\gamma_{min}$ and $\gamma_{max}$.

\begin{figure}
\centering{
\includegraphics[width=65mm]{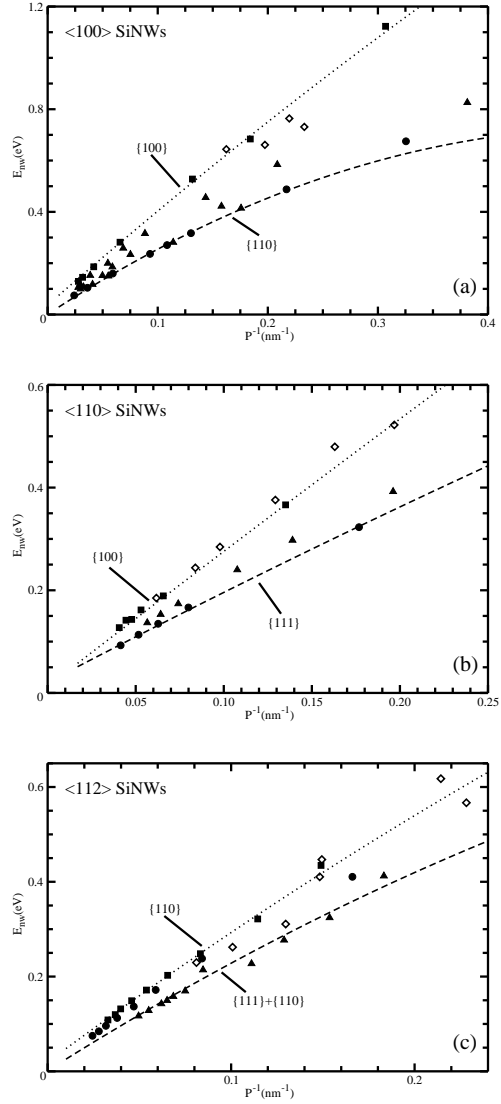}
\caption{Nanowire energy per atom (E$_{\rm nw}$) as function of  
perimeter for (a) $\langle 100\rangle$, (b) $\langle 110\rangle$,
and (c) $\langle 112\rangle$ wires. Close symbols represent 
our results, while open ones the {\it ab initio} data: 
in (a) from ref. \cite{rurali} and (b,c) from ref. \cite{kagimura}.
In (a) the circles ($\bullet$) represent wires with pure $\{110\}$ 
surfaces, the squares ($\blacksquare$) those with pure 
$\{100\}$ surfaces, and the triangles ($\blacktriangle$) those 
with mixed character. A few configurations with mixed character 
fell near the limiting case of wires with pure $\{110\}$ surfaces, 
but they comprised wires with prevailing $\{110\}$ surfaces.
In (b) the circles ($\bullet$) represent wires with prevailing $\{111\}$ 
surfaces, the squares ($\blacksquare$) those with prevailing 
$\{100\}$ surfaces, and the triangles ($\blacktriangle$) 
those with mixed character. In (c) the circles ($\bullet$) represent  
wires with prevailing $\{111\}$ surfaces, the squares ($\blacksquare$) 
those with prevailing $\{110\}$ surfaces, and the 
triangles ($\blacktriangle$) those with mixed character. 
The dotted and dashed lines in (a,b,c) are data fittings  
coming from configurations that determine the energy limits.}
\label{fig2}
}
\end{figure}

Figure \ref{fig2} shows the energy per atom (E$_{\rm nw}$)
of $\langle 100\rangle$, $\langle 110\rangle$, and
$\langle 112\rangle$ SiNWs. This energy is defined with relation to the 
reference crystalline energy per atom, so that for very large wires, 
it tends to zero. We first consider the case of 
$\langle 100\rangle$ SiNWs. Using our classification in terms of wire 
perimeter, the nanowire energies follow a universal scaling law, for 
each facet family. The energy of a nanowire with any surface composition 
(pure or mixed character) falls within a certain region of the graphics, 
always between wires with $\{100\}$ and $\{110\}$ pure surfaces. 
These results would be expected: for a certain wire perimeter, the wire 
energy can have several values, depending on the surface types. 
The crystalline Si $\{100\}$ surfaces have higher energies than  
$\{110\}$ surfaces \cite{bech}, therefore it is consistent that wires 
(with the same perimeter) have higher energies if they have $\{100\}$ 
rather than $\{110\}$ pure surfaces. We performed additional 
calculations, considering wires with other compositions of 
$\{100\}$ and $\{110\}$ surfaces, between those two limiting cases of 
pure surfaces. For those cases, we found the wire energies lying inside 
that region. The figure also shows the results from 
{\it ab initio} investigations \cite{rurali}, that had the 
dimensional parameters renormalized in order
to describe energies in terms of wire perimeter. 
Our results for energies, in terms of interatomic potentials, 
of $\langle 100 \rangle$ SiNWs 
are in good agreement with {\it ab initio} results.

For the $\langle 110\rangle$ SiNWs, we carried simulations for three 
facet families (fig. \ref{fig1}b). Figure \ref{fig2}b 
presents the energies of $\langle 110\rangle$ SiNWs as a 
function of the wire perimeter. The nanowire energies follows an
equivalent scaling law as that of $\langle 100\rangle$ wires. Here, 
the wire energies fall within two limiting curves, which are related to 
wires with prevailing $\{100\}$ surfaces and wires with prevailing 
$\{111\}$ surfaces. The figure also shows the results from 
{\it ab initio} investigations \cite{kagimura}. In \cite{kagimura}, 
SiNWs were constructed with a prevailing cylindrical shape of diameter 
$D$, so that the respective wire perimeters were well characterized by 
$P=\pi D$. Finally, figure \ref{fig2}c presents the energies 
of $\langle 112\rangle$ SiNWs. 
Again, the energies are within two limiting lines, but here one of
the limiting curves is related to nanowires with 
mixed character. According to fig. \ref{fig2}, 
our results are in in good agreement with {\it ab initio} data for all 
three growth directions.

Our results indicate that nanowire 
energies lie within two limiting energy lines, which are directly 
related to the character of the prevailing nanowire surfaces. 
Such results are fully consistent 
with the analytical model presented earlier. For large wire
perimeters, edge effects can be neglected, and there is a linear
relation between energy and the inverse of the wire perimeter.
However, for smaller perimeters, edge effects become 
important \cite{yako,arias},
which could explain the non-linear behavior of that relation. 
It should 
be pointed out that SiNWs with clean reconstructed surface are 
rare \cite{marsen}, while those with hydrogen passivated surfaces are 
more common \cite{ma}. There is a clear scaling 
behavior of wire energies as a function of wire perimeter, no matter if 
the surfaces are clean or passivated. If surfaces were passivated, 
the only difference would be the character of the energy 
limiting lines ($\gamma_{max}$ and $\gamma_{min}$), 
which would be related to the energetics of those passivated surfaces.

\begin{figure}
\centering{
\includegraphics[width=120mm]{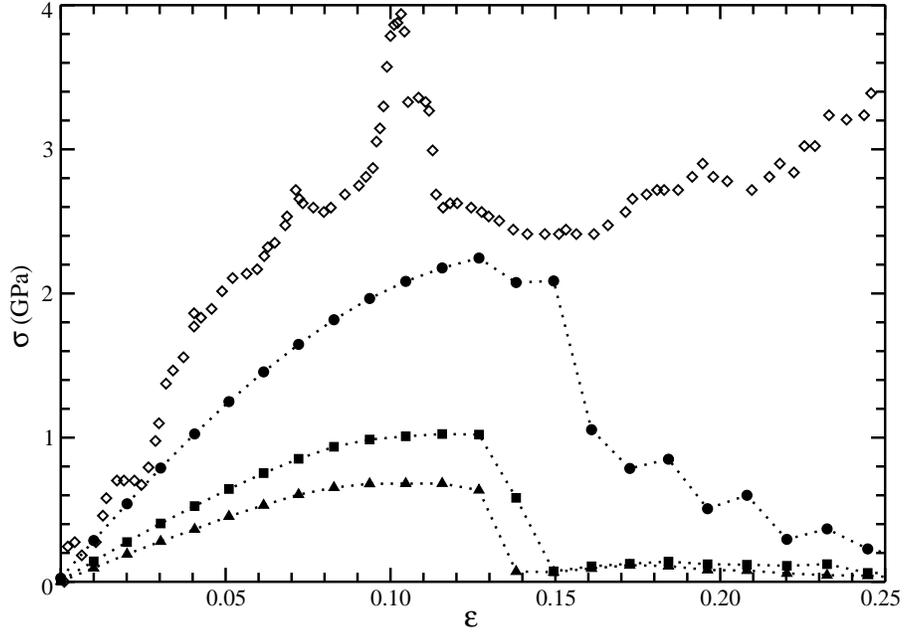}
\vspace*{0.7cm}
\caption{Stress-strain curves as result of tensile deformation processes 
of $\langle 100\rangle$ SiNWs with mixed character 
($\{100\}$ + $\{110\}$). The close (open) symbols represent the results 
of our simulations (experiments). 
The figure shows the results for three different wire 
perimeters: triangles ($\blacktriangle$), squares ($\blacksquare$) and 
circles ($\bullet$) represent respectively wires with 
perimeters of 11.0 nm, 13.2 nm, and 16.9 nm.} 
\label{fig3}
}
\end{figure}

We now discuss the properties of tensile deformation of 
$\langle 100\rangle$ SiNWs, with mixed $\{100\}$+$\{110\}$ 
facets. The deformation simulations  
were performed at constant temperature (350 K), by increasing the 
wire strains, followed by an equilibration process, and then computing the 
resulting uniaxial tensile stresses. Such simulation conditions  
tried to reproduce those of recent experiments on SiNW 
elasticity \cite{kizuka}. Figure \ref{fig3} shows the 
stress-strain relation for SiNWs with several  
perimeters (11.0 nm, 13.2 nm, and 16.9 nm), along with the experimental 
data \cite{kizuka}. For small strains ($\varepsilon < 0.05$), the 
stress-strain curves have a linear behavior, which indicates an elastic 
response. For larger deformations ($0.05 < \varepsilon < 0.13$), 
inelastic behavior takes place. At about $\varepsilon \approx 0.13$, 
there is a large decrease in the stress, for all wire perimeters, that
is consistent with an equivalent behavior observed in another 
theoretical investigation \cite{menon}. The 
stress lowering is observed in experiments only for  
$\varepsilon \approx 0.25$, and indicated the fracture of the 
nanowire \cite{kizuka}. 

The stress-strain curves present a clear trend. For a certain strain, 
the wire stress depends strongly on the nanowire dimensions, 
a phenomenon that could be anticipated considering the dependence 
on the surface/volume ratio. The wire stress is 
the sum of the interatomic forces, normal to a certain cross-section 
area, divided by that area. Atoms in the wire surface are not fully 
coordinated, as those atoms in the wire core. As a result, they give 
only partial contribution to the wire stress. The experimental results, 
presented in the figure, correspond to deformations of ultra-thin
nanowires \cite{kizuka}. Those authors assumed wires with a near 
cylindrical shape, and estimated the smallest wire cross-section of 
about 7.5 nm$^2$, corresponding to a perimeter of about 10 nm. 
Our results on deformation are in good agreement with experiments
for small strains ($\varepsilon < 0.13$). However, the experimental data 
shows that wires
support stronger deformations prior to collapsing. This discrepancy
between theory and experiment may come from limitations of the theoretical
potential in describing highly deformed materials or that the
simulations were not performed in the adiabatic limit. 
On the other hand, according to fig. \ref{fig3},  
the strain of collapse increased with the wire perimeter,  
in the direction of the experimental value. 
However, a direct comparison between theory and
experiments is still difficult, since the wires used in 
those experiments were either hydrogenated or covered by a thin 
oxide layer while ours were reconstructed wires.

\begin{figure}
\vspace*{0.6cm}
\centering{
\includegraphics[width=150mm]{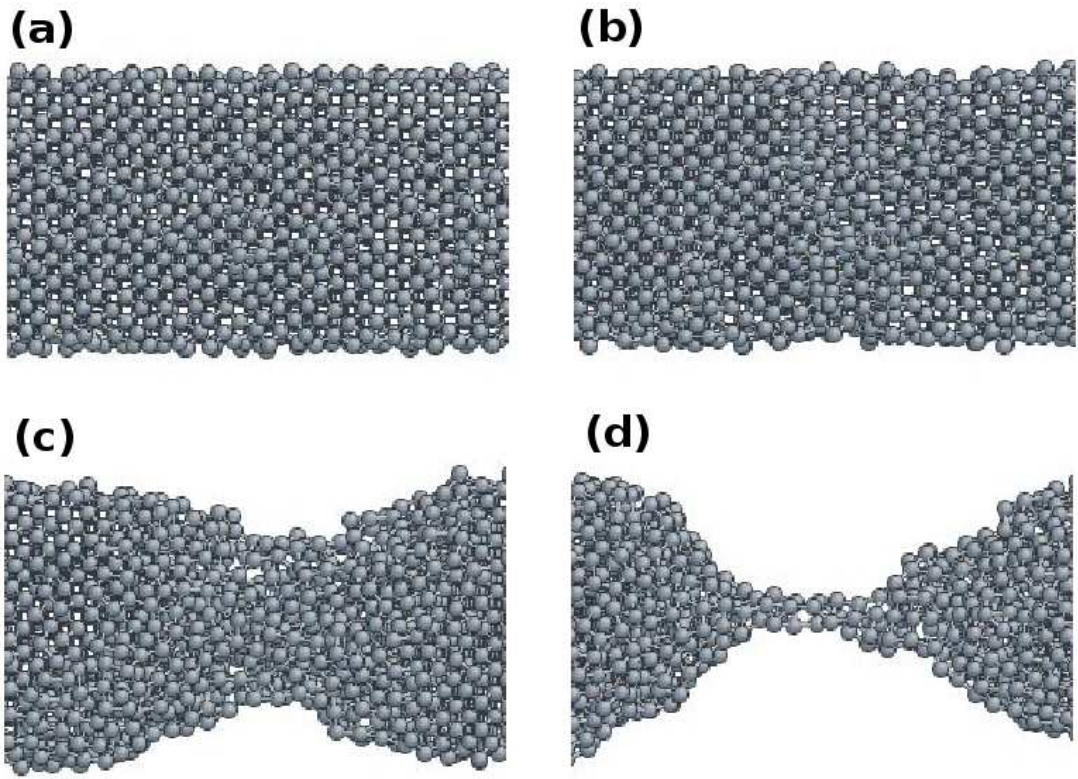}
\caption{Side view of the evolution of a SiNW (with a perimeter of 7 nm) 
as result of external strain. The snap-shots
correspond to configuration under different strains: 
(a) 0.07, (b) 0.08, (c) 0.10, and (d) 0.19.} 
\label{fig4}
}
\end{figure}

An interesting feature emerges from SiNW deformation simulations. 
It would be expected that, for a certain strain, the 
nanowire would simply follow a fracture process, as observed in 
experiments \cite{kizuka}. However, our simulations suggested a 
potentially richer phenomenology. The evolution of the nanowire 
deformation allows larger strains, with nanowire elongations 
considerably larger. The deformation process can be better understood by 
the wire evolution presented in fig. \ref{fig4}. For small strains 
(fig. \ref{fig4}a), the wire only elongates with an elastic response. 
For larger strains (fig. \ref{fig4}b), the wire starts to open a crack 
in the surface. However, instead of this crack just propagating along 
the wire, the wire becomes considerably thinner and continues 
elongating (figs. \ref{fig4}c and \ref{fig4}d). The crack did not 
evolve because atoms in the surface had enough thermal energy to 
diffuse toward the crack, preventing crack propagation.
This mechanism of wire deformation could be useful in creating 
ultra-thin silicon nanowires for several applications, such as 
one-electron transistors. The process could be controlled by an atom 
force microscope operating at a certain temperature, in which the system 
had enough thermal energy to pump surface atoms toward the nanowire crack.
However, this process would require an strict control over the 
applied forces, that we estimate in the order of only a few 
nano-Newtons. 

\section{Summary}

In summary, we found that the perimeter 
is the suitable dimensional parameter to describe the scaling properties
of nanowires. The nanowire energies fall within two 
limiting cases, defined by the prevailing character of 
the wire surfaces. Considering the recent progress in growing
SiNWs, these results provide additional elements to
control the growing of nanowires not only with certain growth directions
but also with certain surface facets. Recent investigations
suggested that the surface facets are important in 
designing nanowires \cite{rurali2}. The surface electronic states
may change the wire electronic properties, that is relevant for 
wire functionalization. For designing nanotransistors, 
one of the current challenges has been how to 
grow and manipulate ultra-thin nanoelements, for
example with specific wire dimensions. 
The relevant effects of quantum confinement, 
and therefore gap engineering, is within a very strict region of wire 
diameters. Our results suggest that nano-mechanical processes, and not 
only chemical ones, may be used to design and create nanowires with 
specific properties.

\acknowledgments

The authors acknowledge support from 
the NANOSENSIM - Nanosensores  Integrados e 
Microsistemas network (contract number 400619/2004-0 CNPq)
and PNM (Programa Nacional de Microeletr\^onica).
JFJ  acknowledges partial support from brazilian agency FAPESP.



\end{document}